\documentclass{vgtc}                          

\ifpdf
  \pdfoutput=1\relax                   
  \usepackage{graphicx}                
  \DeclareGraphicsExtensions{.pdf,.png,.jpg,.jpeg} 
\else
  \usepackage{graphicx}                
  \DeclareGraphicsExtensions{.eps}     
\fi%

\graphicspath{{figures/}{pictures/}{images/}{./}} 

\usepackage{microtype}                 
\PassOptionsToPackage{warn}{textcomp}  
\usepackage{textcomp}                  
\usepackage{mathptmx}                  
\usepackage{times}                     
\usepackage{cite}                      
\usepackage{tabu}                      
\usepackage{booktabs}                  

\onlineid{0}

\vgtccategory{Research}

\vgtcinsertpkg

\title{Labeling of Cultural Heritage Collections\\ on the Intersection of Visual Analytics and Digital Humanities}

\author{Christofer Meinecke\thanks{e-mail: cmeinecke@informatik.uni-leipzig.de}\\ %
        \scriptsize Leipzig University} %

\teaser{
  \includegraphics[width=\linewidth]{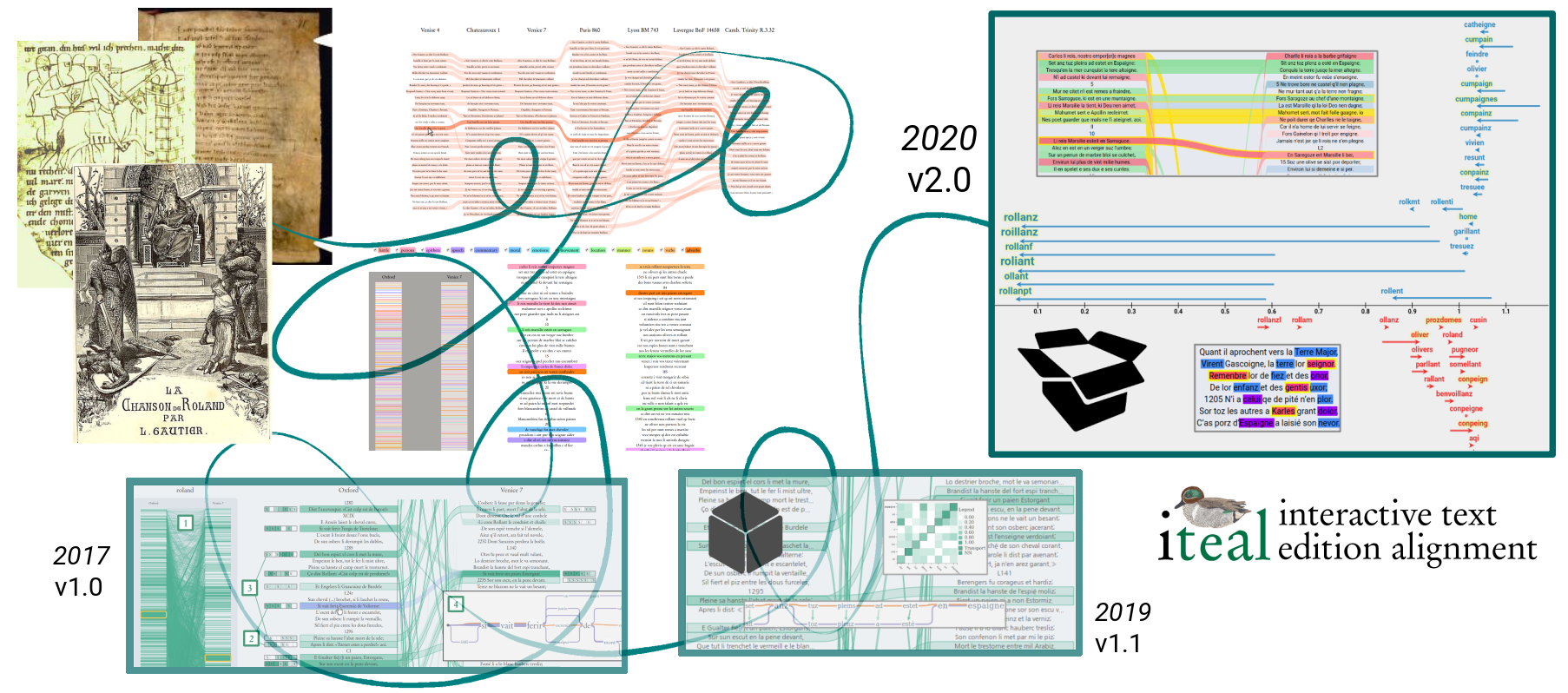}
  \caption{One of the case studies as a speculative process inspired by Hinrichs et al's.~\cite{hinrichs} sandcastle metaphor.}
    \label{fig:teaser}
}

\abstract{
Engaging in interdisciplinary projects on the intersection between visualization and humanities research can be a challenging endeavor.
Challenges can be finding valuable outcomes for both domains, or how to apply state-of-the-art visual analytics methods like supervised machine learning algorithms. 
We discuss these challenges when working with cultural heritage data.
Further, there is a gap in applying these methods to intangible heritage.
To give a reflection on some interdisciplinary projects, we present three case studies focusing on the labeling of cultural heritage collections,  the problems and challenges with the data, the participatory design process, and takeaways for the visualization scholars from these collaborations.
} 


\CCScatlist{
  \CCScatTwelve{Human-centered computing}{Visualization}{};
  \CCScatTwelve{Human-centered computing}{Human computer interaction}{};
  \CCScatTwelve{Applied computing}{Arts and humanities}{};
}

\begin{document}



\firstsection{Introduction}

\maketitle

Cultural heritage data can come in several forms and modalities like text traditions, artworks, music, and crafted objects, or even as intangible like personal biographies of people, performing arts, or cultural customs and rites.
Assigning metadata to such cultural heritage objects is an important task that is done by people working in galleries, libraries, archives, and museums (GLAM) on a daily basis.
These rich metadata collections are used to categorize, structure and study the collections but can be also used to apply computational methods.
For example, in order to apply a supervised machine learning algorithm labels are needed.
The labeling of all objects in a collection is an arduous task when carried out manually because cultural heritage collections contain a large variety of different objects with plenty of details.
Labeling can be seen as either assigning numerical values to an object of interest~\cite{wall2017podium},  assigning categorical labels to an instance~\cite{borghesani2010surfing}, or defining relations between the objects or label categories~\cite{sperrle2019viana}.
All these cases can be either seen as single-label or multi-label problems.
Especially when working with visual material from the GLAM sector multi-label methods are needed in order to assess the variety of depictions on a page or image of interest.
A problem that arises with these collections curated in different institutions is that not always a specific standard is followed and so the used vocabulary can drift apart from another which makes it hard to combine the data from these institutions for large-scale analysis.

We discuss some challenges when working with cultural heritage data. Focusing on the application of machine learning algorithms, intangible heritage data, and valuable outcomes for both communities when engaging in projects on the intersection of visual analytics and the humanities.
Further, we present three case studies focusing on the labeling of cultural heritage collections, with their interdisciplinary team setup, the problems and challenges with the data, the participatory design process, and takeaways for the visualization scholars from these collaborations.
All case studies involve objects of the medieval period and to be more precise, poems in medieval French, and medieval manuscripts and illuminations.
The objectives of the projects range from alignment of textual data, to the visual analysis and exploration of visual corpora, including object detection and the creation of a hierarchy for objects and themes in medieval illuminations.

\newpage
In summary, our contributions are: A discussion of challenges and research gaps on the intersection of digital humanities and visual analytics and a reflection on three case studies for labeling cultural heritage data of the medieval period. 

\section{Challenges with Cultural Heritage Data} 
\begin{figure}[tb]
    \centering
    \includegraphics[width=\linewidth]{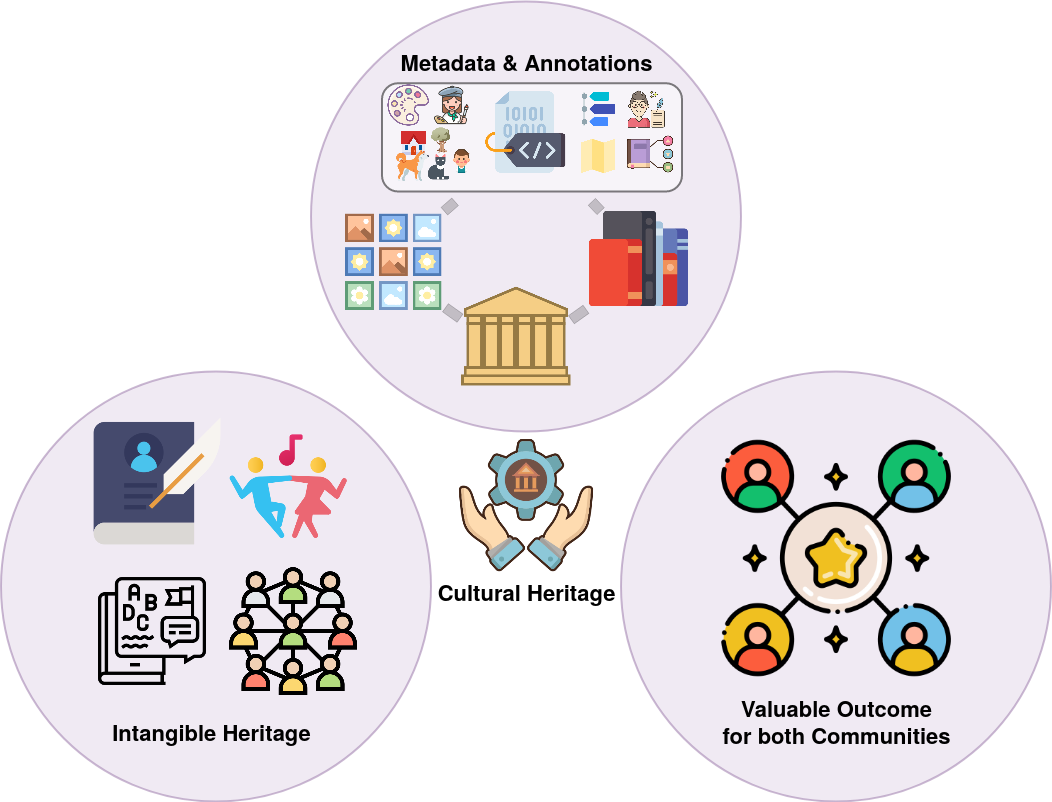}
    \caption{Challenges when working with Cultural Heritage data in interdisciplinary projects.}
    \label{fig:challenges}
\end{figure}

Despite the wider application of machine learning methods on cultural heritage collections~\cite{FIORUCCI2020102}, they are rarely applied together with visualizations to allow further perspectives into the collections.
For collections of visual material, when machine learning methods are applied they are mostly used to project images onto a 2D space~\cite{pfluger2016sifting,crockett2016direct,hochman2013zooming,hristova2016images,yamaoka2011cultural}.
Visual analytics can help in the decision-making process by guiding domain experts through the cultural heritage collection of interest.
Nevertheless, state-of-the-art machine learning methods are often not applicable to the collection of interest because of missing ground truth.
An overview of challenges when engaging in interdisciplinary visualization and cultural heritage projects can be seen in ~\autoref{fig:challenges}. 

\noindent \paragraph{Metadata \& Annotations}

\noindent
The first challenge is the data itself.
Often, only a limited size of data is available or of interest for the domain-specific research question, which makes the application of machine learning methods often not feasible.
Another problem is when there is data but it is not labeled and so no ground truth is available.
In these cases either manual labeling is needed, which can be supported in some cases by visual analytics methods and active learning, or only unsupervised methods can be applied, which again are limited to models trained on other data sources or the data at hand.
In cases where enough data is available and labels and metadata are present the data was often digitized by different institutions leading to a lack of controlled vocabulary between different data sources~\cite{kusnick2020timeline}. 
In these cases, visualization can help in the process of combining the different vocabularies and resolving the inhomogeneities.
A process that otherwise is often done by hand.
Other problems in regards to data quality are incompleteness of the data like missing metadata~\cite{khulusi2019interactive}, damaged material~\cite{khulusi2022exploring}, or imprecision e.g. an OCR approach for text or artifacts in a 3D model. 

When labeling of the data sources is needed in order to conduct an analysis of the data, active learning~\cite{wu2020multi} can be applied in order to reduce the amount of manual labeling.
In this process, a user labels data samples that are queried by an algorithm based on a strategy that optimizes a specific criterion while minimizing the amount of work. 
Furthermore, visualization can help in this process by increasing trust~\cite{chatzimparmpas2020state} and by helping in the sense-making process~\cite{endert2017state}.
For this, Bernard et al. introduced visual interactive labeling (VIAL)~\cite{bernard2018vial} as a concept to combine active learning with visualization systems for the exploration and selection of data points for labeling.
Although the VIAL process gives great results for problems with a small number of classes~\cite{agarwal2018computer,sevastjanova2021questioncomb,ritter2018personalized}, it still was not applied to multi-label problems with a larger number of classes like entities or scenes depicted in visual material of cultural heritage.

\noindent \paragraph{Intangible Heritage}

\noindent
Another open challenge is the analysis of intangible heritage.
Most of the time when cultural heritage data is visualized the focus lies on tangible assets~\cite{windhager2018visualization}.
Only a sparse amount of work focus on intangible assets, like performing arts, crafts, expressions, customs, or rites.
This is even more acute when looking at machine learning methods.
In order to apply machine learning and visual analytics methods to intangible heritage like a specific dance, or oral history, a tangible object like a text, image, or video needs to be created, which then could be enriched by human-annotated labels.
One example of intangible heritage where a tangible asset was created is texts that were passed on orally in vernacular languages until someone wrote them down, like the Song of Roland~\cite{janicke2017visualizing}, which we also include as a case study (\autoref{sec:iteal}). 
Other examples are the comparison of dance movement through video recordings~\cite{arpatzoglou2021dancemoves}, or biographies for prosopographical research~\cite{windhager2017beyond, meinecke2018visual, khulusi2020musixplora}.

\noindent \paragraph{Valuable Outcome for both Communities}

\noindent
The last challenge is not based on the data but the project constellation. Finding valuable outcomes for both the humanities community and the visualization community is not an easy task~\cite{janicke2016valuable}.
Good communication and a participatory design process can help in finding a common research question or several questions that could be of interest to both communities. 
Transparency about the challenges to publish works in the respective fields can also help to prevent conflicts of interest in academic goals and to find a common place to publish the results~\cite{schetinger2019bridging}.
Finding a common vocabulary also helps in bridging the gaps between the different research areas.
Nevertheless, from the visualization perspective, finding ways to evaluate the results is often challenging as no standards exist and quantitative evaluation does not always play a role in humanities research. 
Further, when no ground truth exists for the cultural heritage collection it becomes almost impossible to evaluate the applied methods.

Furthermore, some fields in the humanities are still reserved about using computational methods for their research work.
An example is the domain of art history~\cite{lamqaddam2018tech}.
Drucker said, "To date no research breakthrough has made the field of art history feel its fundamental approaches, tenets of belief, or methods are altered by digital work."~\cite{drucker2013there}
Although this domain has a history of using visualization going back to Aby Warburg's Mnemosyne Atlas in the 19th century such methods are rarely applied\cite{warnke2016prometheus}. 
A way to bridge the gap is to augment the traditional workflows with additional insights, instead of replacing them.
An example is ARIES~\cite{crissaff2017aries} where the operations art historians apply on physical light boxes to curate a collection are imitated in a visualization system.

\section{Case Studies}
Participatory visual design processes~\cite{gappaper} build on, but also extend, task-based development~\cite{munzner2009nested}, as most of the design considerations and adaptations are debated in-depth among all project members~\cite{previs}. 
A side effect of this type of collaboration is the design of visualizations as a speculative process running through multiple iterations by creating several "sandcastle" visualizations~\cite{hinrichs} that help the project discourse.
An example of one of the case studies presented in this paper can be seen as a speculative process in~\autoref{fig:teaser}.
This type of design process also leads to vibrant reflections on required adjustments and gives entirely new visual perspectives on the data at hand. 
Furthermore, it helped in coming up with ways to include domain knowledge as feedback to the visualization system and so to engage in a labeling process of cultural heritage data.
Our projects profited from a longstanding collaboration and trust as a team, which allowed us to avoid major misunderstandings typical of projects at the intersection of visualization and digital humanities~\cite{balancing,drift,cga,schetinger2019bridging}.

\subsection{Interactive Text Edition Alignment}\label{sec:iteal}
The first case study is a project focusing on the alignment of medieval vernacular literature.
To be more precise, the alignment of different versions of the Song of Roland and other works that belong to the genre of French epic poetry~\cite{meinecke2021explaining}.

\noindent \paragraph{Setup}

\begin{figure}[tb]
    \centering
    \includegraphics[width=\linewidth]{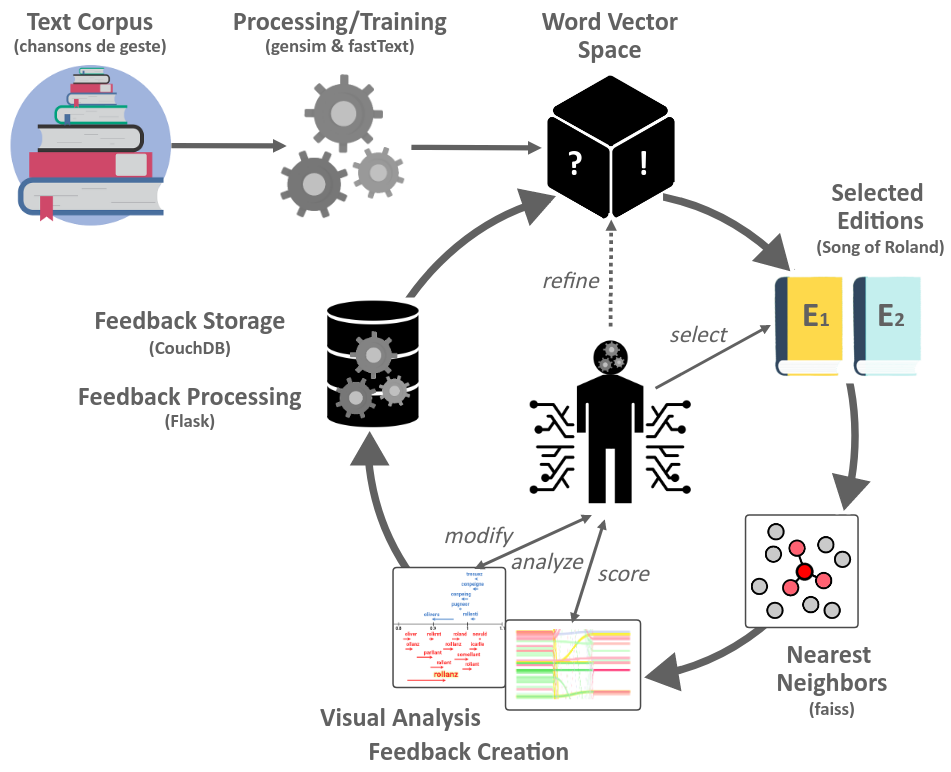}
    \caption{The human-in-the-loop process for Explaining Semi-Supervised Text Alignment through Visualization\cite{meinecke2021explaining}}
    \label{fig:iteal}
\end{figure}

\noindent
For the interactive text edition alignment project, many design iterations were already run through in a participatory design process between one digital humanities scholar and one visualization scholar~\cite{janicke2017interactive, janicke2017visualizing}.
So a common vocabulary was already established when a second visualization scholar joined the project to include methods based on word embeddings to automate the alignment process. 
Still, the new potential methods and how they worked needed to be discussed with all team members over multiple meetings.
After we included these automatic methods new needs to interpret and interact with the alignment were needed~\cite{meinecke2019}. 
An overview of the applied human-in-the-loop process can be seen in \autoref{fig:iteal}. 
The general idea is that a word embedding model is trained on the corpus at hand and then refined by interactions of the domain expert with the words and lines appearing in the poems.

\noindent \paragraph{Data Problems}

\noindent
From a computational side, there were several problems with the data source.
Starting with text artifacts caused by the OCR process, but also including changes in words because of regional and scribal dialect.
The texts of interest were originally orally passed on and later written down in different dialects of medieval french.
Applying word and sentence embedding methods to the data was challenging as there is no pre-trained language model for medieval french and its dialects, which is a common problem for low-resource and under-resourced languages.
Furthermore, the whole alignment process is highly interpretive, which becomes a problem when evaluating the method, as no ground truth for an alignment tuple is available.
The only way to evaluate the methods was to present the domain expert with two alignments to rate, one before domain knowledge was induced and one after, without them knowing which alignment was which.

\noindent \paragraph{Design Process}

\noindent
To include methods to interpret and interact with the alignment we engaged in a participatory design process by meeting once a week to discuss potential visualization designs.
The first challenge was to give reasons to the domain expert why a particular alignment occurred.
For this, we first introduced a heatmap showing the similarities of the word vectors and highlighting their nearest neighbor and transportation pairs for the computation of the Word Movers Distance~\cite{kusner2015word}.
With these new means to understand why an alignment occurred, came also the need to accept or reject alignments, but because of the highly interpretive nature of the poetry at hand, we decided to use a Likert scale to label the alignments and so to induce the domain knowledge into the alignments.
The general idea was that the underlying word vector distribution should change based on the rating. 
For this, we adapted the Rocchio algorithm~\cite{rocchio1971relevance}.
The problem with the Likert scale approach was that half-line alignments would get a low score, so alignments where one line is split into two lines in another version. 
A reason for this is the usage of different meters in the poems.
Because of this, we applied a binning approach to treat the half-line alignments differently.
The new interaction method of scoring the alignments created the need to see changes in the alignment after an interaction i.e. which alignments are added and which are removed.
Further, more direct and easily understandable methods to change the word vectors were needed as well as ways to see the changes in the word vector neighborhood.
For this, we added methods to change the distance between word vectors based on drag and drop, which can be seen as labeling the relation of the words with a numerical value.
Then, we also added methods to see the changes in the neighborhood of word vectors over multiple iterations.
In order to better communicate which parts of the poem and which words were strongly affected through the interactions, we added word-level heatmaps showing either how strong a word vector or its neighborhood has changed.

\noindent \paragraph{Takeaways}

\noindent
The usage of the system showed that in the beginning the labeling of word relations was rarely used, but in becoming more familiar with the system the domain expert used this labeling method primarily in the end.
We conclude that easy interaction methods for labeling with direct feedback such as moving a word from one position to another position are more appealing than more complex ones that are not easy to grasp, such as applying a scoring method to multiple words and sentence components.
Also, complex views to compare the parameter-based approach with the automatic one were not needed, the same goes for views that show the distribution of the similarity values through the iterations.
It is important to note that we needed to align as a team before we could engage in the textual alignment of medieval poetry.

\subsection{Visualizing Entities in Medieval Manuscripts}

\begin{figure}[tb]
    \centering
    \includegraphics[width=\linewidth]{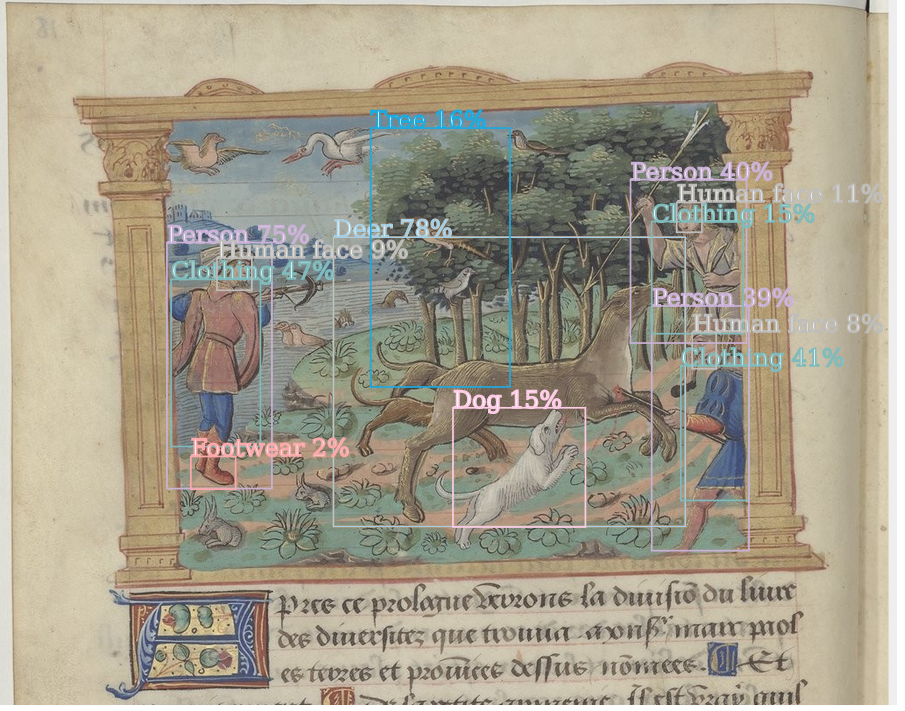}
    \caption{A page of the Marco Polo dataset with entities found by a neural network.}
    \label{fig:dh2022}
\end{figure}
The second case study is about visualizing entities in medieval manuscripts~\cite{dh2022}.
The idea was to focus on the similarities of images in manuscripts in a similar way as we did the textual alignment.
For this, we applied computer vision methods and visualization as a starting point for distant viewing on a small dataset that is composed of 700 medieval illuminations of the French Marco Polo textual tradition.

\noindent \paragraph{Setup}

\noindent
The project constellation was the same as in the first case study, two visualization scholars with experience in several digital humanities projects, and one digital humanities scholar/medievalist interested in medieval manuscripts. 

\noindent \paragraph{Data Problems}

\noindent
In the beginning, we applied object detection with a Faster R-CNN~\cite{ren2015faster} trained on ImageNet~\cite{imagenet_cvpr09} for feature extraction and Open Images~\cite{kuznetsova2018open} for the detection.
The problem was that these datasets and their underlying hierarchies do not match the entities depicted in medieval illuminations.
This regards the contemporary vocabulary used in the hierarchy like "airplane" or "television", and also to the depiction of the entities~\cite{hall2015cross}.
Furthermore, the dataset was not annotated with bounding boxes, we did not have a list of object classes of interest, and the dataset was not big enough to train a new network,
But still, the network trained on natural images provided a first impression and a convenient starting point for creating new classes and extracting some initial training data for the classes that are appropriate for the domain and the period.

\noindent \paragraph{Design Process}

\noindent
In order to analyze the results, we build a visual interface for the exploration of the classes. 
For this, it was important to allow browsing the different classes and to compare all depictions with a specific class in a visualization.
Further, the domain expert could select contemporary classes to be removed.
On an image level, it was important to allow to show all detection results as bounding boxes with their confidence score and to allow filtering based on classes and confidence to prevent visual clutter.
An example page with detected objects can be seen in \autoref{fig:dh2022}.
For a specific bounding box, it was also possible to display the most similar bounding boxes e.g. to see the most similar Human Faces to a specific Human Face in the dataset.
Also, the possibilities to draw new bounding boxes, create new classes, and relabel existing bounding boxes were needed.

\noindent \paragraph{Takeaways} 

\noindent
The project leads us to think about other larger visual corpora including artworks~\cite{vm} and Paris Bibles (\autoref{sec:paris}). 
It also showed us that there is a need for label hierarchies with period-specific terminology and also for methodologies to unify the vocabulary of different datasets.
Furthermore, manual labeling of these datasets takes a lot of time so there is a need to integrate either visual interactive labeling for multi-label problems or weak supervision~\cite{inoue2018cross} to reduce the amount of manual work.

\subsection{Hierarchical Classification for Medieval Illuminations}\label{sec:paris}
\begin{figure*}[tb]
    \centering
    \includegraphics[width=\linewidth]{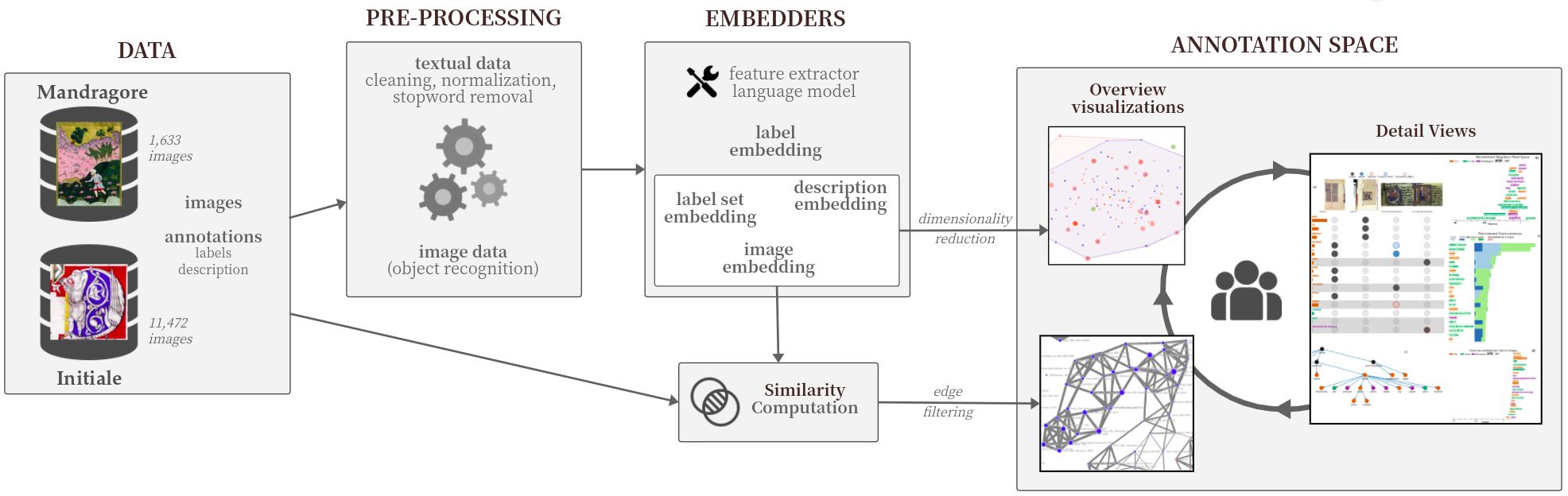}
    \caption{Systematic overview of the semi-automated image labeling workflow applied to the Paris Bible datasets~\cite{parisbible}.}
    \label{fig:paris}
\end{figure*}

The last case study is about combining two different multi-modal datasets of very similar material~\cite{parisbible}.
The goal was to create a common label vocabulary of two datasets of medieval illuminations that belong to different institutions and are annotated with different types of metadata.
Also to create a label hierarchy that is appropriate based on domain and time-specific properties for the classification and detection of entities inside the illuminations.
A systematic overview of the workflow can be seen in \autoref{fig:paris}.
A future goal is to investigate how the visual depictions changed based on their geospatial and temporal context, and also to study their similarities and differences.

\noindent \paragraph{Setup}

\noindent
The project constellation from the other case studies was extended by one medievalist interested in medieval literature and manuscripts. 

\noindent \paragraph{Data Problems}

\noindent
One problem is the automatic extraction of the illuminations from the manuscript pages.
Each digital image illustrates one or two pages of a specific manuscript and can contain one or several illuminations depicting various scenes and objects. 
This process is hard to standardize because of the different quality of the background material, as well as some illuminations or ornaments can take up almost the whole page.

Another problem lies in the vocabulary that was used for the annotations.
Both datasets were created at different times and priorities.
Still, the vocabulary of both datasets was initially based on the "Thésaurus Garnier"~\cite{garnier1984thesaurus}, but both institutions deviated from this controlled vocabulary.
This divide between the datasets is based on the history of the different institutions and is a major problem when scholars would like to see relationships in the larger picture of cultural heritage data.
Furthermore, similar images from the same unit of text have completely different attention paid to them, while some images have no or only a few annotations others have over 50 different annotations. 
The inconsistencies in the datasets are also based on different decisions of several annotators and an annotation process that range over several years.

\noindent \paragraph{Design Process}

\noindent
We started with creating exploratory visualizations for the different facets of the dataset.
Including presenting the images on a timeline, image clouds, set visualization of annotation combinations, comparing the annotation of different manuscripts, and a faceted search based on annotations and metadata.
But all these methods were not as helpful as we imagined at this stage of the project.
The reason for this was the inconsistencies in the annotation and metadata.
So we started to focus on grouping the images first by manuscripts and different similarity measurements. The result was a graph visualization that helped in selecting subsets of manuscripts based on different criteria.
From there a point cloud showed the images projected to a two-dimensional space based on text or image features.
This helped in selecting a subset of images to enrich their labels simultaneously and to build a hierarchy of the vocabulary.
We decided to focus on labeling multiple similar images simultaneously to help detect similar depicted themes and objects that are not used consistently over both datasets, thus also helping in bringing the datasets together. The recommendation also helps as the vocabulary of the metadata is rather large and not all terms are known by the domain experts.
This is probably a similar problem the original annotators of the dataset had, as the labeling is not always consistent in the data source.

\noindent \paragraph{Takeaways}

\noindent
This research project led to several ideas for future research direction in terms of human-in-the-loop scenarios for image classification and object detection, as well as the comparison of different graphs and hierarchies putting the labels in context.
Nevertheless, a paper about the project was initially rejected at a high-quality visualization venue, with the recommendation to submit the "Greatly written and didactically prepared paper" to a domain-specific venue.
The major problem lies in a missing evaluation with a baseline for labeling, which is hard to do as most systems do not tackle the problem of multi-label classification at the scale of several thousand possible classes. 
Even state-of-the-art approaches like VIAL were not applied prior to this type of data.
A paper about this topic may have a higher impact in a domain-specific venue, but we still had new ideas based on this initial rejection about possible research directions that could be of interest to the visualization community, that can be evaluated and also could help the domain experts in the final goal of analyzing the illuminations.

\section{Conclusion}
We discussed challenges when dealing with cultural heritage data, addressing issues of data size, metadata, quality of the data, and intangible heritage.
As well as the current lack of projects in the GLAM sector focusing on multi-label classification by applying human-in-the-loop processes like visual interactive labeling in order to bridge the gap between machine learning, digital humanities research, and visualization.
Further, we presented three case studies of close collaboration with (digital) humanities scholars engaging in a participatory design process focusing on the labeling of tangible cultural heritage.

\acknowledgments{
The author wishes to thank Estelle Guéville, David Joseph Wrisley, and Stefan Jänicke for fruitful discussions and the insights gained from the previous and ongoing collaborations. 
}

\bibliographystyle{abbrv-doi}

\bibliography{template}
\end{document}